*Research paper*
# Misplaced trust? The relationship between trust, ability to identify commercially influenced results, and search engine preference

Sebastian Schultheiß[1] and Dirk Lewandowski[2]

[1] Hamburg University of Applied Sciences, Germany; sebastian.schultheiss@haw-hamburg.de
[2] Hamburg University of Applied Sciences, Germany; dirk.lewandowski@haw-hamburg.de

## Abstract

People have a high level of trust in search engines, especially Google, but only limited knowledge of them, as numerous studies have shown. This leads to the question: To what extent is this trust justified considering the lack of familiarity among users with how search engines work and the business models they are founded on? We assume that trust in Google, search engine preferences, and knowledge of result types are interrelated. To examine this assumption, we conducted a representative online survey with n = 2,012 German internet users. We show that users with little search engine knowledge are more likely to trust and use Google than users with more knowledge. A contradiction revealed itself — users strongly trust Google, yet they are unable to adequately evaluate search results. This may be problematic since it can potentially affect knowledge acquisition. Consequently, there is a need to promote user information literacy to create a more solid foundation for user trust in search engines.

## 1. Introduction

Search engines enjoy a good reputation and a high degree of trust among users, as numerous user studies [e.g., 1, 2] and surveys [e.g., 3] have shown. Users also rate their ability to use search engines as (very) high [e.g., 4]. We must however assume this attitude is not based on competent evaluation, since search engine users display a marked lack of knowledge regarding search engine functionality [5, 6] and business models [4]. Hence, it can be assumed that user understanding of increasingly complex SERPs is inadequate, and that users are unaware of their lack of knowledge. This can lead to clicks that are not the result of an informed decision, for instance, when a user clicks on an advertisement assuming it is an organic result.

Current search engine result pages (SERPs) exhibit a very complex structure. They contain organic results, ads (text ads, shopping ads), verticals, direct answers, and knowledge graph results [4]. The search engine market is dominated by Google, which holds a market share of 87% in the United States [7] and 93% in Europe [8] across all platforms as of August 2020. Google generates its revenues mainly through ads (search-based advertising). In 2019, 83% of Google's annual revenue of $162 billion was generated through advertising [9]. Ads currently receive a black 'ad' label, with the design of the label having changed repeatedly over the years, becoming more subtle in the process [10].

Keyword-related advertising, also referred to as paid search marketing (PSM), belongs to the field of search engine marketing (SEM). SEM also includes search engine optimization

(SEO), which is defined as 'the practice of optimizing web pages in a way that improves their ranking in the organic search results' [11]. While external actors strive to make their content more visible through PSM or SEO activities, search engines such as Google are also pursuing their own interests. There is potential for these interests to be reflected in rankings. This influence may also become apparent within Universal Search results, i.e., blended SERPs composed of different verticals such as products, maps, and video. In this case, Google is not only an intermediary for external services, but also a provider of its own services (e.g., Google Shopping, YouTube) and therefore competes with offerings from other companies. In the case of Google Shopping, the European Commission found that Google was giving its own comparison shopping service an illegal advantage and consequently fined Google €2.42 billion [12].

Thus, on current SERPs, numerous actors can be identified who are capable of influencing the visibility of results. These actors can be categorized in the stakeholder groups presented in Figure 1. Those who influence the SERP in addition to search engine operators include search engine optimizers (SEOs), paid search marketers (PSMs), and the users themselves, whereby the latter are also influenced by the SERP.

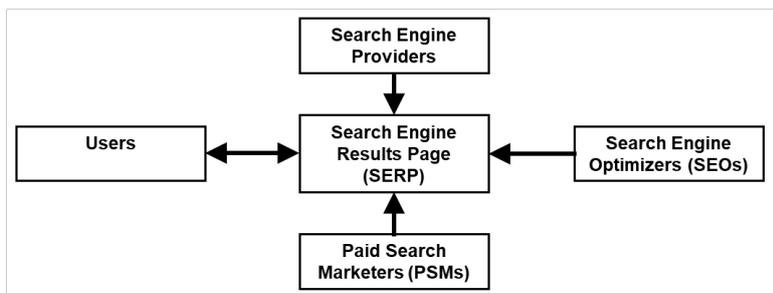

**Figure 1: Stakeholder groups that influence results**

In this paper, we focus on PSM and SEO. In the following, we refer to 'SEO results' and 'PSM results,' where the former refers to organic, non-paid results and the latter to ads (see also section 'Marking tasks' for the assignment of search results to PSM and SEO). We assume correlations exist between the user's ability to distinguish search results based on potential influence by PSM or SEO activities (in the following referred to as 'skill'), the user's trust in Google, and the user's search engine preferences. We assume that a high level of trust in Google is associated with users who have never (sufficiently) explored the inner workings of search engines and the structure of SERPs, leading to difficulties when they attempt to identify the areas affected by PSM and SEO. We also suspect these users mainly use Google (often perceived as the 'standard search engine' due to its dominant position in the market) and see no reason to use anything else. Accordingly, we assume that a low level of trust in Google is accompanied by a more critical view of (and more skill in using) search engines in general (see Figure 2).

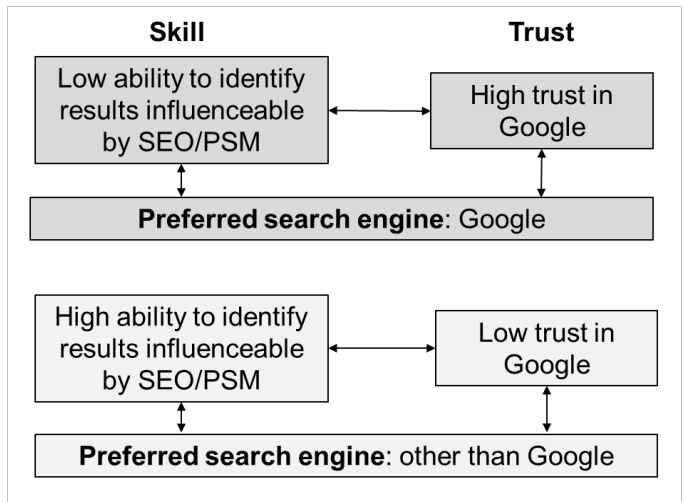

Figure 2: Expected correlations between skill, trust, and search engine preference (top: users with a low level of skill; bottom: users with a high level of skill)

Our objective is to investigate the extent to which trust in Google correlates with (1) ability to differentiate result types and (2) search engine preference as well as (3) how the ability to differentiate among result types relates to search engine preference. To the best of our knowledge, this has not yet been investigated. This is of importance since a high level of trust accompanied by a low level of knowledge regarding result types could lead to undiscriminating search engine use (of Google in particular) and thus to a possible influence on information acquisition. To examine these correlations, we conducted a representative online survey with n = 2,012 participants.

The rest of the paper is structured as follows. First, we provide an overview of studies on user trust in search engines and search engine user information literacy. Then, we describe the methods we used. After describing how we analyzed the data, we provide the results in six parts. We present the results of the survey separately, e.g., questions about trust in Google, as well as correlations with other results, e.g., search engine preferences. Finally, we discuss the results, provide a summary, and present suggestions for future research.

## 2. Literature review

### 2.1 User trust in search engines

Search engine users have a high level of trust in search engines and Google in particular. This has been shown in laboratory studies and (online) surveys. Laboratory studies show that users choose top results even if they are less relevant [1, 2] or less credible [13] than results displayed lower in a ranked list. Unkel & Haas [14] confirmed these findings. They manipulated the credibility cues *reputation*, *neutrality*, and *social recommendations* of the results and showed that only *reputation* had an influence on selection behavior. However, this influence was again surpassed by the effect of the results position. Thus, search engine users follow Google's results ranking more than their own assessments or other influencing factors. This implies a high level of trust that search engines will rank the 'best results' at the top of the result list.

In addition to the findings from laboratory studies, users also explicitly stated in representative surveys that they trust information found by using a search engine. In a representative survey in the United States, three quarters of respondents said they trusted the information they found on search engines. 28% do so for all or almost all, 45% for most information [3]. Search engine trustworthiness is comparable to traditional news media, as shown by a representative study of Internet users from 28 markets including the United States, China, and Germany [15]. A study with representative samples from countries in Europe, the Americas, and the Asia-Pacific region plus South Africa examined the trust in information found in search engines in the context of news. 33% of participants stated they mostly trust news they find through search engines. The trust in news found on social media channels is 23%. It should be noted, however, that only 42% of all respondents said they trust news in general and that there are strong differences between the surveyed countries (trust in news in Finland is 59%, but only 24% in France) [16].

Similarly, surveys without representative samples show a high level of trust in search engines among users. An online survey of students by Westerwick [17] revealed that a high result ranking increases sponsor credibility, which in turn affected the credibility of the message. Thus, user trust in search engines influences the perceived credibility of the displayed search results. In a (non-representative) online survey with students by Klein et al. [18] roughly 40% of respondents assessed the information found on search engines as mostly reliable, while around 5% considered it to be 'always reliable.' These rates place search engines somewhere between libraries and databases (both of which were considered by most respondents to be mostly or always reliable) and internet communities, forums, blogs, and podcasts (which were considered not reliable or only somewhat reliable). In a survey of college students, Taylor & Dalal [19] found evidence of gender differences in the assessment of trustworthiness. The survey results indicate that men were more likely than women to trust search engines to provide objective results. In an exit survey following an experiment, Ramos and Eickhoff [20] found that the subject's trust in the result list increased when the results were accompanied by explanations of how the individual search terms contributed to the rank.

## 2.2 Search engine user information literacy

A representative survey of internet users in the United States revealed a high self-assessment of search engine users regarding their search skills. Most users (91%) stated they usually find what they are looking for, while 56% were very and 37% somewhat confident in their search abilities [3]. In practice, however, a different picture emerges, as numerous user studies show.

The subjects of a study by Stark et al. [5] had problems formulating precise queries and exhibited insufficient knowledge of the ranking criteria used by search engines. The problem of formulating succinct queries can also be observed among children. Dragovic, Madrazo Azpiazu, & Pera [21] suggested a solution to this problem involving a module that creates queries to capture the information needs children try to express. A study by Singer, Norbisrath, & Lewandowski [6] showed that search engine users had difficulties especially when solving complex search tasks. The subjects estimated their success rates in solving complex tasks to be much higher than they actually were. The contradiction between self-assessed and actual information literacy was also noted by Lewandowski et al. [4]. The survey included questions about Google's business model, marking tasks for ads, and questions on how the users self-assess their own research skills. It was shown that search engine users often have little or no knowledge of how search engines generate income and cannot reliably distinguish ads from organic results. At the same time, however, most users (90.8%) reported that they had 'excellent' or even 'perfect' (the two highest values on a six-point scale) research skills.

Users tend to overestimate their information literacy not only in the context of commercial search engines, but elsewhere, too. Mahmood [22] compiled a literature review of 53 studies on the self-assessment of information literacy published between 1986 and 2015, without further limitation to search engines or specialized databases. In most of the studies, self-assessment exceeded the actual information literacy of participants. Similar findings were made by Douglas et al. [23] and Holman [24] for first-year students, Ngo, Pickard, & Walton [25] for upper secondary students (in both cases using questionnaires), as well as by Cullen, Clark, & Esson for junior physicians searching in medical databases [26].

It can be summarized that with commercial search engines, as with other information systems, the high self-assessments of users do not correspond to actual information literacy.

## 3. Research questions

As described in the introduction, we aim to investigate the relationships between user trust in Google, the ability to differentiate results influenceable by SEO or PSM, and search engine preference. In order to consider all possible relations, the research questions are as follows:

**RQ1**: How does user trust in Google relate to the user's ability to differentiate SEO and PSM results?

**RQ2**: How does user trust in Google relate to search engine preferences?

**RQ3**: How does user ability to differentiate SEO and PSM results relate to search engine preferences?

## 4. Materials and methods

We conducted a representative online survey with n = 2,012 German internet users. The online survey was conducted as part of the SEO-Effekt[1] (SEO Effect) project, which has the goal of describing and explaining the role of SEO from the perspectives of the participating stakeholder groups. The complete online survey includes sections that go beyond the focus of this paper. In this paper, the results of the questionnaire sections that deal with search engine use, trust, and ability to assign results influenceable by SEO/PSM are presented. Other sections, such as personalization or further usage habits, will not be considered. The survey was carried out together with the market research company 'Fittkau & Maaß Consulting'[2] (hereinafter abbreviated by F&M) between January and April 2020. The subjects were recruited by the online panel provider 'respondi'[3] in cooperation with F&M. We had no direct contact to the subjects. The market research company (F&M), which carried out the survey in cooperation with us, operates according to the principles of the UN Global Compact. This means that F&M operates in a way that fulfills fundamental values regarding human rights, labor, environment, and anti-corruption. Written consent to process their data was obtained from all participants. Data was analyzed anonymously.

---

[1] https://searchstudies.org/seo-effekt/
[2] https://www.fittkaumaass.de/

## 4.1 Sampling

We used a sample that is representative of the German online population, in accordance with the criteria specified by 'Arbeitsgemeinschaft Onlineforschung' (Working Group Online Research; [27]). The population includes German internet users ranging in age from 16 to 69. From the total sample of n = 2,012 subjects, two sub-samples of n = 999 subjects (large screen) and n = 1,013 subjects (small screen) were formed, which meet the same requirements regarding representativeness described above. Sample 1 attended the survey with a large screen (e.g., desktop PC, laptop, tablet; 'large screen' group), sample 2 with a small screen (smartphones; 'small screen' group). We consider search on small screens in addition to desktop search, as mobile search now exceeds desktop search in terms of search volume [28]. The subjects were invited by e-mail and each received €0.75 for completing the survey.

## 4.2 Questionnaire development

The survey had three sections. These are 'usage,' 'trust,' and 'ability to assign results influenceable by SEO/PSM.' We additionally collected demographic data. As can be seen in the questionnaire in Table 1, we based the questions on two thematically related studies [3, 29], one being on search engine use in the United States, the other investigating German users' understanding of PSM. Question 1.1 on search engine preference is designed to distinguish between Google users and users of other search engines in the analysis. Questions 2.1.1 and 2.1.2 have been derived from Purcell et al.'s study [3] with slight adjustments in question formulation and response options. They aim to determine the user's trust in Google and the information found through its search engine. Finally, questions 3.1-3.4 served to find out how accurately the respondents can assign search results to the potential influences (SEO, PSM). For further details on the marking tasks, see also section 'Marking tasks'.

**Table 1. Questionnaire**

| No. | Question | Ref. |
|---|---|---|
| I) Usage | | |
| 1.1 | 'Which search engine do you use most often?' Response options: Bing, Ecosia, DuckDuckGo, Google, Web.de, Yahoo!, Another, I don't know/not specified | [3] |
| II) Trust | | |
| 2.1 | 'If you think of Google: To what extent do you think the following statements apply to Google?' 2.1.1 Google is a fair and unbiased source of information 2.1.2 The information I find through Google is correct and trustworthy Response options: absolutely correct[1], correct[1], neutral[2], rather not true[2], doesn't apply at all[2], I don't know | [3] |
| | Information part 'SEO/PSM' | |
| III) Ability to assign results influenceable by SEO/PSM | | |
| 3.1 | 'You will now see a Google result page. Are there any search results on this page that can be influenced by the website operator paying Google?' | [29] |
| 3.2 | 'One more question about this search result page: Are there any search results on this page that can be influenced by search engine optimization?' | |

---

[3] https://www.respondi.com/EN/

| No. | Question | Ref. |
|---|---|---|
| 3.3 | 'You will now see another Google result page. Are there any search results on this page that can be influenced by the website operator paying Google?' | [29] |
| 3.4 | 'One more question about this search result page: Are there any search results on this page that can be influenced by search engine optimization?' | |
| | Response options 3.1-3.4: Marking the requested results or skipping the task by specifying that the requested result type is not available on SERP. | |

[1] agreement
[2] disagreement

All questions are closed questions, including single response questions, rating scale questions, and questions with marking options for SERP screenshots. For comparison purposes (see results section) we divided the response options of questions 2.1.1 and 2.1.2 into 'agreement' and 'disagreement.' We classified 'absolutely correct' and 'correct' as 'agreement,' the other response options as 'disagreement.' We classified the response option 'neutral' as 'disagreement,' since in our view this option implies a certain degree of skepticism regarding the trustworthiness of Google. Before section III of the questionnaire, we integrated the following explanations of SEO and PSM to ensure that all respondents could understand the marking tasks in the survey:

'Website operators have several ways to ensure that their web pages appear at the top of the Google result page for a specific query, namely I) Payment: They pay money to Google[4], or II) Search engine optimization: They design their websites accordingly, e.g. by using certain keywords, quick page speed, and appropriate image titles and descriptions. Next, we will show you two different Google result pages and would like to ask you whether or which results can be influenced by payment to Google and/or search engine optimization.'

The survey was conducted in German. See Appendix 1 for the original questionnaire in German.

## 4.3 Marking tasks

For the marking tasks, we created two blocks with a total of four tasks. Tasks A and B were assigned to block I (simple), tasks C and D to block II (complex). Two blocks were created to address a variety of SERP elements and to differentiate between basic and complex SERPs. The structure of the two SERPs per block is identical in terms of the elements on the SERP. The queries and elements of the SERPs are shown in Table 2.

Table 2. Marking tasks: queries and SERP elements

| Block | Task | Query (translated) | SERP elements |
|---|---|---|---|
| I (simple) | A | tax return help | Organic results, Text ads (top and bottom of SERP) |
| | B | legal advice | |
| II (complex) | C | apple iphone | Organic results, Text ads (top of SERP), Shopping ads, News, Knowledge Graph |
| | D | samsung galaxy | |

---

[4] This is a simplified explanation. The fact that a payment to Google is only made after an ad is selected was left unmentioned for the sake of comprehensibility.

Since we tested all tasks on two devices (large and small screen), we created a total of eight SERP screenshots. Each participant received two SERP screenshots, one randomly assigned from block I and one randomly assigned from block II (e.g., SERP screenshots for tasks A and D). Each SERP was shown two times. First, all PSM results were to be marked (i.e. paid results) and second, all SEO results (i.e. unpaid search results that are influenceable by SEO measures). As shown in Table 3, there are differences in the number of shopping ads and news results between screen sizes for tasks C and D. The reason we kept these differences is that we considered a realistic representation of the SERPs more important than an exact matching of the set of search results between both screen sizes.

**Table 3. Marking tasks: results to be marked**

| Task | Device | Area | Results to be marked |
|---|---|---|---|
| A | Large and small screen | SEO | Organic results (10*) |
| A | Large and small screen | PSM | Text ads, top of SERP (2*) Text ads, bottom of SERP (2*) |
| B | Large and small screen | SEO | Organic results (10*) |
| B | Large and small screen | PSM | Text ads, top of SERP (2*) Text ads, bottom of SERP (2*) |
| C | Large screen | SEO | Organic results (6*) News (3*) |
| C | Large screen | PSM | Text ads, top of SERP (2*) Shopping ads (8*) |
| C | Small screen | SEO | Organic results (6*) News (2*) |
| C | Small screen | PSM | Text ads, top of SERP (2*) Shopping ads (2*) |
| D | Large screen | SEO | Organic results (6*) News (3*) |
| D | Large screen | PSM | Text ads, top of SERP (2*) Shopping ads (8*) |
| D | Small screen | SEO | Organic results (6*) News (2*) |
| D | Small screen | PSM | Text ads, top of SERP (2*) Shopping ads (2*) |

The screenshots were created using the desktop version of the Chrome browser. We used browser extensions for window resizing [30], user-agent switching [31] and screen capture [32] to create screenshots for all SERPs in both display formats ('large screen' and 'small screen'). We used version 2.10.14 of the graphics software GIMP [33] to crop the SERP screenshots down to the elements we wanted to investigate as shown in Table 3. We also matched the small-screen SERP screenshots with the large-screen SERP screenshots in terms of results and their positions. Otherwise, different selection behavior in the survey might not be attributable to the screen size (large vs. small), but to somewhat different results in certain positions. See the research data for the SERP screenshots. In Figure 3, we show a section of the large screen SERP of task A ('tax return help'). The PSM results are indicated by a dashed line, the SEO results by a solid line.

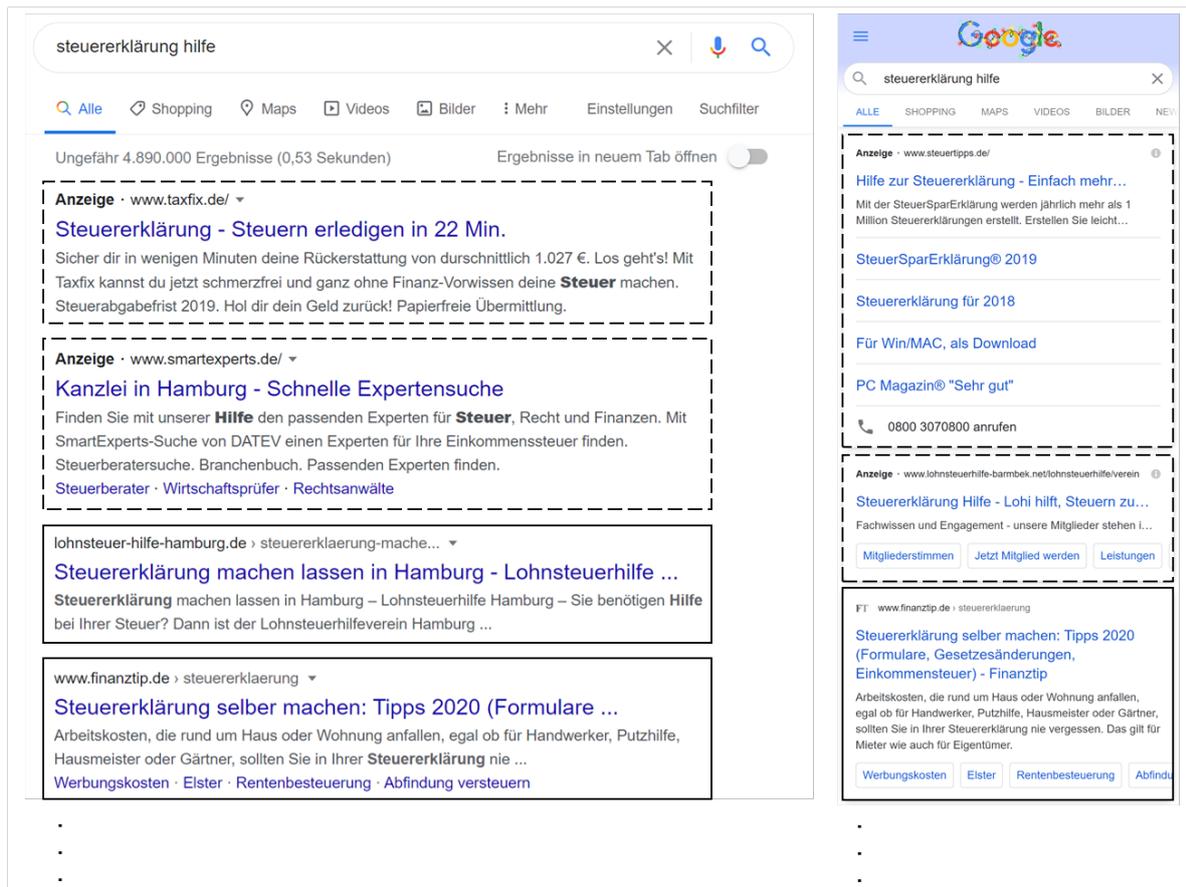

Figure 3: SERP for task A, large screen (PSM results: dashed line, SEO results: solid line)

## 4.4 Data analysis

### 4.4.1 Success rates for marking tasks

Based on the marked elements, a success rate was calculated for each participant per task (A-D), device (large, small), and area (SEO, PSM). This rate accounts for correctly marked (true positive) and incorrectly marked (false positive) results using the formula $\frac{\text{n true - n false}}{\text{n to be marked}}$.

Two examples follow, the first for achieving a positive success rate for task A, large screen, SEO results. In this case, 10 organic results are to be marked, of which the subject marks 8 results (8 true). In addition, the subject incorrectly marks 2 ads (2 false). This results in a success rate of 0.6. Negative success rates are also possible, if a subject makes more incorrect than correct markings, exemplified by task B, small screen, PSM results. In this case, a total of 4 text ads are to be marked. If a subject identifies all 4 text ads (true), but additionally marks 6 organic results (false), the subject achieves a success rate of -0.5.

### 4.4.2 Statistical methods

To calculate the correlation between trust in Google and success rates, we performed Spearman's rho analyses, as this is an appropriate method for a correlation between ordinal and metric data. Chi-square tests of independence including the calculation of effect sizes were performed to examine the relation between trust in Google and the preferred search engine. For the comparison of success rates with different search engine preferences (Google vs. non-Google users; for sample sizes see section 'Characteristics and search engine preference of internet users'), we conducted Mann-Whitney $U$-tests as the data did not meet the requirements of t-tests in terms of normal distribution. The alpha level for all statistical tests was set to .05. We used IBM SPSS Statistics version 27 for data analysis [34].

## 5. Results

We describe the results in six parts. First, major demographic characteristics of the subjects and their search engine preferences are reported. Then, the success rates for the identification of PSM and SEO results are presented. Next, we show the trust users have in Google. Finally, we examine the relationships between trust in Google, success rates, and search engine preferences.

### 5.1 Characteristics and search engine preference of internet users

The online survey included n = 2,012 subjects, of which 51.6% were male and 48.4% female. In addition to age distribution and highest level of education, Table 4 also shows which search engines the subjects prefer. The vast majority of users prefer Google (89.8%). Thus, only few users prefer other search engines such as Ecosia (5.6%) or DuckDuckGo (2.0%).

Table 4: Participant characteristics

| | % | n |
|---|---|---|
| **Gender** | | |
| Female | 48.4% | 974 |
| Male | 51.6% | 1,038 |
| **Age** | | |
| under 16 | 0.0% | 0 |
| 16 to 24 | 15.6% | 313 |
| 25 to 34 | 19.7% | 397 |
| 35 to 44 | 20.1% | 404 |
| 45 to 54 | 23.5% | 472 |
| 55 to 69 | 21.2% | 426 |
| 70 and older | 0.0% | 0 |
| **Highest level of education** | | |
| Certificate of Secondary Education without completed apprenticeship | 2.3% | 46 |
| Certificate of Secondary Education with completed apprenticeship | 10.2% | 206 |
| General Certificate of Secondary Education | 29.6% | 595 |
| A-levels | 29.7% | 597 |
| University degree | 24.3% | 488 |
| None | 0.1% | 1 |
| (Still) without school-leaving certificate (e.g., student) | 1.2% | 24 |
| Other | 2.7% | 55 |
| **Preferred search engine** | | |
| Google users | 89.8% | 1,790 |
| non-Google users | 10.1% | 201 |
| *Ecosia* | *5.6%* | *111* |
| *DuckDuckGo* | *2.0%* | *39* |
| *Bing* | *1.6%* | *32* |
| *Web.de* | *0.5%* | *10* |
| *Yahoo!* | *0.3%* | *5* |
| *Another* | *0.2%* | *4* |
| I don't know/not specified | 0.1% | 1 |

### 5.2 Success at identifying SEO/PSM results

In this section, we present results on the ability of subjects to correctly label results influenceable by SEO or PSM. The corresponding questions are 3.1-3.4. The success rates were calculated as described in section 'Success rates for marking tasks'.

In Figure 4, within each bar, the average success rate is shown, for example SERP A on large screen and the marking of PSM results, 0.628. Above the bar, the number of participants to whom this success rate refers is given. In the case of SERP A, large screen, PSM, this means 80.3% of those tasked with evaluating SERP A on a large screen made markings for PSM results while 19.7% incorrectly indicated that there were no PSM results on SERP A.

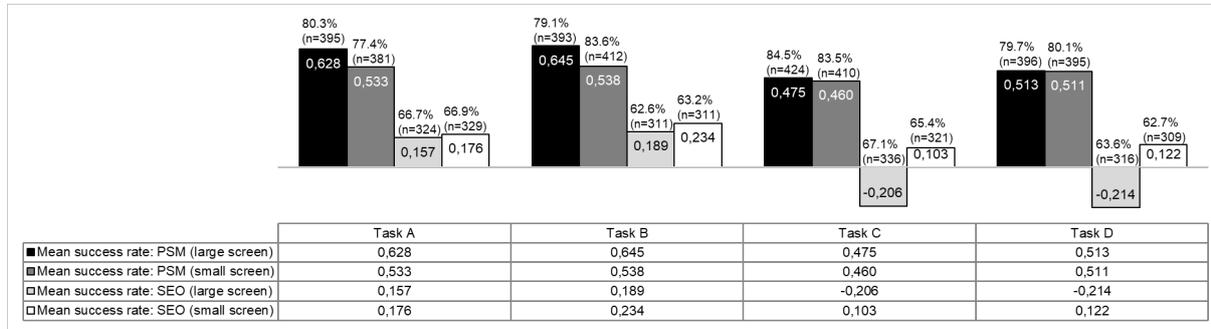

Figure 4: Success rates for the identification of SEO/PSM results

When looking at the bars for PSM and SEO results, PSM results were identified much better than SEO results on all SERPs. In addition, both result types were identified better on the simply structured SERPs (A and B) than on the more complex SERPs (C and D). Also, we can see that subjects made markings for PSM results (up to 84.5%) considerably more often than they made markings for SEO results (up to 67.1%).

When comparing both screen sizes, PSM results were better identified on the large screen than on the small screen for the simple structured SERPs (A and B). On the large screen, the negative success rates of the complex SERPs (C and D) for the SEO results show that, on average, more false than correct markings were made. For the complex SERPs (C and D), SEO results were identified with much greater accuracy on the small screen than on the large screen. It should be noted here that the calculation of the success rates includes the false positives (in this case the ad markings). On the large screen of the complex SERPs, eight shopping ads were shown, on the small screen, only two. Thus, the success rate decreases much more when a respondent incorrectly marks all eight shopping ads on the large screen as SEO results than when the subject incorrectly marks the (only) two shopping ads on the small screen as SEO results.

## 5.3 Trust in Google

In the following, the findings from the questions on trust in Google are presented (Table 5). First, we asked whether Google is a fair and unbiased source of information. The option 'neutral' is selected most often (38.4%), followed by 'correct' (25.5%) and 'rather not true' (17.1%). Second, the subjects were asked if they consider the information found in Google to be correct and trustworthy. The most frequent answer again was 'neutral' (44.3%), followed by 'correct' (37.0%). Thus, user opinions of Google's trustworthiness are more positive than negative.

**Table 5. Trust in Google**

|  | Google is a fair and unbiased source of information (2.1.1) | The information I find through Google is correct and trustworthy (2.1.2) |
|---|---|---|
| absolutely correct | 8.7% | 9.0% |
| correct | 25.5% | 37.0% |
| neutral | 38.4% | 44.3% |
| rather not true | 17.1% | 6.0% |
| doesn't apply at all | 6.5% | 1.3% |
| I don't know | 3.7% | 2.4% |

## 5.4 Correlation between trust in Google and success rates

In the following, we examine whether the statements on trust in Google and success rates are related to one another. For the scores, those respondents who indicated that the result type they were looking for is not found on the corresponding SERP are not considered, since these respondents did not make any markings.

We find significant positive correlations between the questions on trust in Google and the mean success rates for PSM and SEO results with small effect sizes (Table 6). For classifying effect sizes as small, medium, or large, see [35]. However, due to the large sample sizes tested, we can assume stable correlations [36]. In other words, a low level of trust in Google is correlated with better performance at identifying PSM and SEO results.

**Table 6. Spearman's rank correlation coefficients between trust in Google and success rates**

|  |  |  | Trust in Google | |
|---|---|---|---|---|
| Question No. |  |  | 2.1.1 | 2.1.2 |
| Spearman's rho | Success rate PSM (mean) | Correlation Coefficient | .122** | .041* |
|  |  | Sig. (1-tailed) | $p < .001$ | $p = .046$ |
|  |  | N | 1,642 | 1,666 |
|  | Success rate SEO (mean) | Correlation Coefficient | .199** | .137** |
|  |  | Sig. (1-tailed) | $p < .001$ | $p < .001$ |
|  |  | N | 1,400 | 1,417 |

\* $p < .05$

\*\* $p < .001$

## 5.5 Comparison of trust in Google with search engine preferences

We assessed whether subjects who prefer to use Google are more likely to trust in Google than subjects who prefer other search engines. For this we used the grouped response options 'agreement' and 'disagreement' as shown in the questionnaire in section 'Questionnaire development'.

The relation between the variables 'trust in Google' and 'preferred search engine' are significant for both questions on trust in Google. For question 2.1.1 (Google as a fair and unbiased source of information), there is a significant relationship between the variables ($\chi^2(1, N = 687) = 33.753, p < .001$), with a small effect size ($\varphi = .132, p < .001$). For question 2.1.2 (Information found through Google is correct and trustworthy) a significant relation with a small effect size can be observed again ($\chi^2(1, N = 924) = 43.203, p < .001$), $\varphi = .148, p < .001$.

The differences are also illustrated in Figure 5. Accordingly (and unsurprisingly), users who prefer to use Google trust Google more than non-Google users.

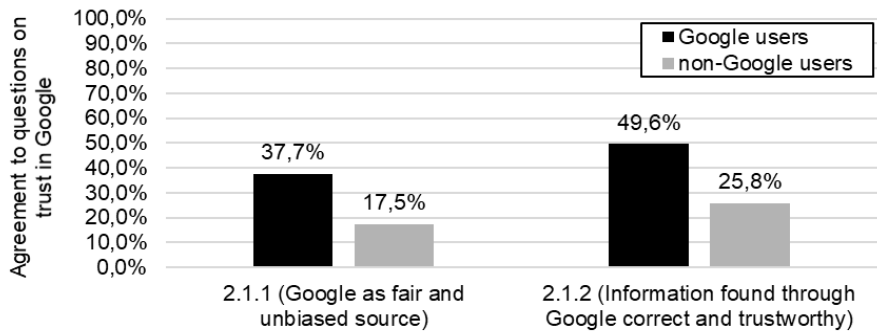

**Figure 5: Agreement to questions on trust by search engine preference**

## 5.6 Comparison of success rates with different search engine preferences

In the following, we examine whether respondents who prefer Google are more (or less) successful than respondents who prefer other search engines. Mann-Whitney *U*-tests were carried out to determine whether the differences are significant. Subjects who stated that they mostly used Google achieved significantly lower success rates in the identification of PSM and SEO results than subjects who mostly use other search engines, as Table 7 shows.

**Table 7. Success rates by search engine preference**

|  | PSM | | SEO | |
| --- | --- | --- | --- | --- |
|  | Google users | non-Google users | Google users | non-Google users |
| success rate (mean) | .513106 | .561441 | .091563 | .166369 |
| success rate (SD) | .504271 | .543263 | .440147 | .464146 |
| *p* values (Mann-Whitney *U*-tests) | .041 | | .022 | |

## 6. Discussion

Regarding the task of identifying search results that could be influenced by SEO or PSM, the PSM results were identified much better than the SEO results. Markings were also made more frequently (whether correct or incorrect) when the task involved identifying PSM results. This suggests that Internet users are generally more familiar with the concept of paid results than they are with SEO. In addition, for SERPs with complex structures, correct responses for both result types were considerably less frequent than for SERPs with a simple structure. The respondents have a rather positive image of Google's trustworthiness. Among those respondents who expressed a low level of trust in Google, a better identification performance of PSM and SEO results was found. The same subjects (with low trust in Google) are more likely to use alternative search engines. The group of non-Google users also achieves a better identification performance of PSM and SEO results. It can thus be concluded that (1) a low level of trust in Google correlates with better identification performance as well as the use of search engines other than Google, and, inversely, (2) a higher level of trust in Google correlates with lower identification performance and preferred use of Google.

One way to characterize the first group mentioned above (low trust, high skill, non-Google) is that this could be users who at least have a basic understanding of search engines and their result types, are therefore more critical towards Google, and are more often motivated to prefer other search engines. Accordingly, the second group (high trust, low skill, Google) could be users who, due to their lack of knowledge, see no reason to mistrust Google and therefore do not feel the need to use a different search engine.

The study shows that Internet users have difficulties in identifying PSM and SEO results on complex SERPs and SEO results in general. The subjects apparently had difficulties associating organic results with possible SEO influences and distinguishing between them and PSM results. These findings are in line with the studies on low information literacy [e.g., 4–6] and interviews with experts (Authors, 2020). The generally positive opinion on the trustworthiness of Google as shown by other studies [e.g., 3] was also confirmed.

The PSM results were identified with greater accuracy on the large screen than on the small screen. A possible explanation is the better distinguishability of the result types on the large screen, since organic results on the small screen are provided with icons at the same position where the ads had their ad label, which could further complicate differentiation from ads.

The study has limitations. First, the sample is limited to German Internet users. Second, the design we used does not allow us to draw conclusions on causal relationships. This presents an opportunity for future studies to follow up on our results. In an experiment, for instance, a subset of subjects could be provided in advance with information about the result types in order to investigate whether varying levels of knowledge are the cause of the different trustworthiness assessments observed.

Our findings result in a problem for those search engine users who have a high level of trust in search engines, but who cannot sufficiently identify and differentiate among result types. The users with high trust also represent the largest user group. Thus, there is a need to raise awareness about the different actors on search result pages in order to create an understanding of the different interests of the actors on search result pages, e.g., by means of tutorials, scaffolding, or gamification as described by Karatassis [37]. Promoting search engine literacy is crucial for putting the knowledge acquisition of search engine users on a solid foundation. However, in the context of information literacy, search engine literacy is hardly considered [38]. Based on our findings, we argue that information literacy training must start where users search in their everyday lives, namely with commercial search engines. If users recognize their skill deficits based on search engines, they should also be more open to more advanced training, such as on how to select and use specific databases.

## 7. Conclusions

We investigated the extent to which the generally high level of user trust in Google correlates with skill and preferred search engines. For this purpose, we conducted an online survey with n = 2,012 participants representative of German Internet users between the ages of 16 and 69. We find that users who have little knowledge of the results types are more likely to trust and favor Google than users with greater knowledge. This suggests a rather unsuspecting attitude towards Google search results, which has the potential to affect the knowledge acquisition of the users. We therefore consider it necessary to promote the information literacy of search engine users. This could either relativize the high level of trust in search engines or bring it on a more solid basis in terms of the various influences on modern search engine result pages.

## 8. Research data

## 9. Acknowledgements

## 10. Funding

This work is funded by the German Research Foundation (DFG - Deutsche Forschungsgemeinschaft), [grant number 417552432].

# 12. Appendix 1: Questionnaire (German)

**Table 8. Questionnaire (German)**

| Nr. | Fragestellung | Quelle |
|---|---|---|
| I) Nutzungsverhalten | | |
| 1.1 | 'Welche Suchmaschine nutzen Sie am häufigsten?' Antwortmöglichkeiten: Bing, Ecosia, DuckDuckGo, Google, Web.de, Yahoo!, Eine andere, weiß ich nicht/keine Angabe | [3] |
| II) Vertrauen | | |
| 2.1 | 'Wenn Sie an Google denken: Wie sehr treffen die folgenden Aussagen Ihrer Meinung nach auf Google zu?' 2.1.1 Google ist eine faire und unvoreingenommene Informationsquelle 2.1.2 Die Informationen, die ich über Google finde, sind korrekt und vertrauenswürdig Antwortmöglichkeiten: trifft voll und ganz zu[1], trifft zu[1], neutral[2], trifft eher nicht zu[2], trifft gar nicht zu[2], weiß ich nicht | [3] |
| | Informationsteil 'SEO/PSM' | |
| III) Fähigkeit, Ergebnisse zuzuordnen, die durch SEO/PSM beeinflussbar sind | | |
| 3.1 | 'Kommen wir nun zu der ersten Google-Ergebnisseite. Existieren auf dieser Seite Suchergebnisse, auf die Einfluss genommen werden kann, indem Google dafür vom Website-Betreiber bezahlt wird?' | [29] |
| 3.2 | 'Noch eine weitere Frage zu dieser Suchergebnisseite: Gibt es hierauf auch Suchergebnisse, auf die mit Hilfe von Suchmaschinenoptimierung Einfluss genommen werden kann?' | |
| 3.3 | 'Kommen wir nun zu unseren Fragen zur zweiten (und letzten) Google-Ergebnisseite. Existieren auf dieser Seite Suchergebnisse, auf die Einfluss genommen werden kann, indem Google dafür vom Website-Betreiber bezahlt wird?' | [29] |
| 3.4 | 'Noch eine weitere Frage zu dieser Suchergebnisseite: Gibt es hierauf auch Suchergebnisse, auf die mit Hilfe von Suchmaschinenoptimierung Einfluss genommen werden kann?' | |
| | Antwortmöglichkeiten 3.1-3.4: Markierung der jeweils gefragten Suchergebnisse oder Überspringen der Aufgabe, indem angegeben wird, dass der gewünschte Ergebnistyp sich nicht auf der SERP wiederfindet. | |

[1] Zustimmung

[2] Ablehnung